\documentclass[preprint,showpacs,preprintnumbers,amsmath,amssymb]{revtex4}

\def\be{\begin{equation}}
\def\ee{\end{equation}}
\def\bea{\begin{eqnarray}}
\def\eea{\end{eqnarray}}

\def\ssc{\scriptscriptstyle}
\def\lsim{\mathrel{\raise.3ex\hbox{$<$\kern-.75em\lower1ex\hbox{$\sim$}}} }
\def\gsim{\mathrel{\raise.3ex\hbox{$>$\kern-.75em\lower1ex\hbox{$\sim$}}} }

\begin{document}

\preprint{{\vbox{\hbox{NCU-HEP-k028}
\hbox{Apr 2007}\hbox{rev. Oct 2007}
}}}

\vspace*{1.5in}

\title{The Quantum World is an {\boldmath\protect  AdS$_5$} with the
Quantum Relativity Symmetry {\small \boldmath\protect  $SO(2,4)$}}

\author{\bf Otto C. W. Kong
%$^1$ and $^2$
%\vspace*{.2in}
}
%\email{otto@phy.ncu.edu.tw}

\affiliation {Department of Physics and Center for Mathematics and
Theoretical Physics, National Central University, Chung-li, Taiwan 32054  \\
%$^2$Department of Physics, University of , USA \\
}

\begin{abstract}
Quantum relativity as a generalized, or rather
deformed, version of Einstein relativity with a 
linear realization on a classical six-geometry beyond the 
familiar setting of space-time offer a new framework to
think about the quantum space-time structure.
% It also gives some picture of a quantum (noncommutative) 4D space-time. 
The formulation requires two deformations to be implemented through 
imposing two fundamental invariants. We take them to be
%present a proposition that the two quantities are 
the independent Planck mass and Planck length. Together, they
gives the quantum $\hbar$. The scheme leads
to {\small \boldmath\protect $SO(2,4)$} as the relativity symmetry. The quantum 
world has an AdS$_5$ `classical' geometry, which is parallel 
to the ``conformal universe", but not scale invariant. 
%The results do not depend on any other theoretical
%assumption, dynamical or otherwise. We discuss some physics picture
%of the formulation.
\end{abstract}

\pacs{ }
\maketitle

\noindent{\em Introduction :- \ }
This letter is in a way a sequel of Ref.\cite{023}, in which an interesting new
perspective to think about the subject of Quantum Relativity has been introduced.
Since the early days of the quantum theory, the notion about some plausible
`quantum structure' of space-time itself at the microscopic limit, in connection
to quantum gravity or otherwise, has been pondered by many theorists.
The analysis of Ref.\cite{023} leads to a radical proposal for the description
of space-time at the microscopic or quantum level --- that it has to be described
as part of something bigger, here we dub the  `quantum world', with  
dimensions beyond space and time. Authors of Ref.\cite{023}
wanted to present a detailed and clear account of theoretical formulation and
thinking behind the radical proposal. We hope this letter can bring that
perspective to the attention of a much broader audience.

On the other hand, we do have a very important conceptual
breakthrough, at least within the subject domain of Quantum
Relativity, to be presented here. It leads to an important new result
--- the identification of the quantum world as an AdS$_5$ with
{\small \boldmath\protect  $SO(2,4)$} as the relativity symmetry.
Readers familiar with some of the more popular themes in the
literature going under the title of AdS-CFT
correspondence~\cite{adscft} and the holographic
principle~\cite{holo} may incline to consider  
the particular result as what has been well  known.  The
 {\small \boldmath\protect  $SO(2,4)$} group algebra has been discussed as
the symmetry of `quantum gravity' in six dimensions.
%, or specifically an AdS$_5$ space-time. 
The relevance of {\small \boldmath\protect
$SO(2,4)$} and  AdS$_5$ to quantum gravity obtained from the two
very different approaches may suggest both are likely to be getting
something right about Nature. However, the connection really ends
there. The two approaches are logically completely independent. The
study of  AdS$_5$ gravity is basically taking AdS$_5$ as the
space-time, or rather part of an AdS$_5\times \rm{S}_5$ compactification
of the ten dimensional superstring theory. One can justify the
introduction of  AdS$_5$ somewhat independent of the `string'
assumption, based on the closely related perspective of Yang-Mills
theory at the  large-$N$ limit. Our approach here, however, has
nothing to do with any of that. While many theorists may believe in
string theory or  large-$N$ Yang-Mills theory to be useful in
describing Nature, one should bear in mind that we have no
experimental evidence of that. String theory still has to produce an
experimental verifiable prediction; and the Yang-Mills theories that
have been established experimentally have only $N\leq 3$. 
Our approach, arguably, starts with a much more minimal set of theoretical assumptions. 
We look for a direct description of quantum physics (of the classical notion of
space-time), as for example also adopted in Ref.\cite{A}. This is in
contrast to the `quantization perspective', adopted for instance in
most studies within the string theory framework, within which one
finds a `classical' fundamental description and produce the quantum
counterpart through some quantization procedure. Ref.\cite{A} starts
with a new, supposed to be more fundamental dynamic principle,
assuming a (matrix model) noncommutative geometry. We start with a new relativity
symmetry and obtain a dual classical and noncommutative geometric
description, before any consideration of dynamics. One should also note that
works along our perspective is only at a very preliminary stage. We
believe it deserves a good chance to be seriously considered and
developed.  %-4

%Relativity is about reference frame transformation symmetry.
%Inverting the point of view, the relativity symmetry can be used to
%construct the basic arena on which we think about putting up the
%reference frames --- {\it e.g.} the Minkowski space-time of Einstein relativity. 
Physicists know of two relativities, namely Galilean
relativity of Newtonian physics and Einstein relativity. Interesting
enough, going from the former to the latter can be considered a
direct result of the mathematical procedure of symmetry deformation
or stabilization~\cite{spha}. Such stabilizations may be considered
as the only legitimate symmetries to describe physics as confirming
an unstable symmetry to be the `correct' symmetry requires infinite
experimental precision, establishing $1/{c^2}$ as
exactly zero in this case. The same procedure
applied again (to the Poincar\'e symmetry) with minimal physics
inputs leads a new relativity. Taking notion from quantum physics,
such as the existence of a fundamental Planck scale, leads to a new
relativity to be considered as the Quantum Relativity. The
approach  is actually technically quite
simple and considered to be accessible to a broad range of
physicists beyond that of the `high energy theorists'. %-7

%\section
\noindent{\em Quantum relativity :-\ }
To get a better idea of what symmetry deformation is all about, let us
take a look at the Galilean to Einstein case. Think about the algebra of the
generators for Galilean boosts. Deforming the zero commutators of any
two Galilean boosts to $1/c^2$ times a corresponding rotation
generator gives the Lorentzian {\small \boldmath\protect  $SO(1,3)$}
symmetry. If there has to be a velocity with  magnitude $c$ invariant
under reference frame transformations, the above is the unavoidable
mathematical consequence. The right physics interpretation 
says that we should think about 4D  (Minkowski) space-time instead of 3D
space as the basic arena for fundamental physics with Lorentzian, or
Poincar\'e, symmetry.

The idea of a quantum relativity dates back more than half a
century~\cite{S}. A simple way to put it is to say that if quantum
physics introduces the idea of the Planck scale, one may want it to
be characterized by a reference frame independent quantity. For
instance, you do not want to see the Planck length to suffer from a
Lorentz contraction. It has been realized that that can only be done
by modifying, or rather deforming, the relativity symmetry,
basically in the same way as deforming the Galilean {\small
\boldmath\protect $ISO(3)$} algebra to the Lorentzian {\small
\boldmath\protect $SO(1,3)$}. The first symmetry for  such a quantum
relativity suggested was essentially {\small \boldmath\protect
$SO(1,4)$}~\cite{CO}, though some recent authors bringing back this
old topic preferred to think about it outside the Lie algebra
framework~\cite{dsr,tsr}. Sticking to the Lie algebra
deformation perspective~\cite{spha}, Ref.\cite{023} gives a very
radical but otherwise sensible physics picture of the quantum
relativity through a linear realization. There, the quantum
relativity symmetry was identified as {\small \boldmath\protect
$SO(1,5)$}, through one further deformations as suggested in so-called triply
deformed special relativity \cite{tsr}. The three deformations, from
the Galilean relativity, are parametrized by the speed of light $c$,
the Planck scale as a mass-energy scale $\kappa c$ as originated in
Ref.\cite{S}, and a sort of infrared (``length") bound associated with 
the cosmological constant ~\cite{tsr}. The radical physics picture follows 
naturally the lesson from Einstein. Just like the 3D space is part of the 4D 
space-time, the 4D space-time is then part of the quantum world --- a
(classical) 6D geometry with the two extra dimensions being
something beyond space and time!

Here in this letter, we present rather {\small \boldmath\protect
$SO(2,4)$} as the symmetry for the quantum relativity. It is still a
three deformation setting, but the last deformation is done
differently. The latter is still implemented through a limiting
length, but rather on the ultra-violet. It is the Planck length,
$\ell$. The  starting point is the important
observation  that implementing a Planck mass is not enough
to get the relativity quantum, because there is no $\hbar$. In fact,
interpreting Planck energy and Planck length as essentially one (the
Planck) scale assumes quantum physics. Formulating a quantum
relativity should rather be trying to get that as a result. We have
to produce explicitly the $\hbar$ as an invariant~! The current
letter present exactly such a formulation, getting $\hbar$ as
$\kappa c\ell$. This is breaking with the unquestioned tradition
since 1947!  An AdS$_5$ hypersurface within the six-geometry is
obtained as the quantum world. We will also discuss some physics
features of the new {\small \boldmath\protect $SO(2,4)$} relativity.

%\section
%\noindent{\em The three deformations to {\small \boldmath $SO(2,4)$}:-\ } 
We have discussed briefly the deformation of the Galilean
{\small \boldmath $ISO(3)$ the {\small \boldmath $SO(1,3)$ Lorentz symmetry. 
%What is deformed is the part of the algebra for the
%Galilean boosts. Such (velocity) boosts are originally spatial
%translations dependent on an external parameter --- time, {\it i.e.
%$\Delta x=v\,t$}. The deformation turns the boosts into space-time
%rotations (Lorentz boosts) as the {\small \boldmath $SO(1,3)$ is
%linearly realized on the Minkowski 4D space-time. The mathematical
%Lie algebra deformation admits also {\small \boldmath $SO(4)$} as  a
%solution. It is the physics consideration that identifies {\small
%\boldmath $SO(1,3)$ with the deformation parameter $1/c^2$ taken
%such that $c$ is a limiting velocity under the Newtonian picture. %-9
The mathematics of further deformations are basically the same. We
skip most of the explicit details, but summarize in Table~1 the
essential aspects of the three deformations. Note that $\eta_{\ssc
\mathcal M\mathcal N} =( 1, -1, -1, -1, -1, 1)$ with the indices go
from $0$ to $5$; $\eta_{\ssc A\!B}$ is the $0$ to $4$ part; other
than that, it is the usual notation. $J_{\ssc \mathcal M\mathcal N}$
here denotes the 15 generators of the the {\small \boldmath
$SO(2,4)$} algebra accordingly.

%%%%%%%%%%%%%%%%%%%%%%%%%%%%%%%%%%%%%%%%%%%%%%%%%%%%%%
\begin{table}[b]
 \caption{The Three Deformations Summarized:-}
%\footnotesize
\begin{center}
\begin{tabular}{|c|c|c|}    \hline\hline
$\Delta x^i(t) = {v^i} \cdot t$   &   $\Delta x^\mu(\sigma) =
{p^\mu} \cdot \sigma$
 &  $\Delta x^{\ssc A}(\rho) = {z^{\ssc A}} \cdot \rho$  \\
\hline
{$|v^i|\leq c$}           &   {$|p^\mu |\leq \kappa\,c$ }           &    {$| z^{\ssc A} |\leq i\, \ell $} \\
$- \eta_{ij} v^i v^j = c^2 \left(1-\frac{1}{\gamma^2}\right) $
 &   $\eta_{\mu\!\nu} p^\mu p^\nu = \kappa^2 c^2  \left(1-\frac{1}{\Gamma^2}\right)$
      & $\eta_{\ssc \!A\!B} z^{\!\ssc A} z^{\!\ssc B}= - \ell^2 \left(1+\frac{1}{G^2}\right)$     \\
\hline
$M_{{\ssc 0}i}\equiv N_i \sim P_i$                  &   $J_{\mu\ssc 4}\equiv O_\mu \sim P_\mu $
      &   $J_{\ssc \!A\!5}\equiv O'_{\ssc \!A} \sim P_{\ssc \!A} $    \\
{$[N_i, N_j]  \longrightarrow -i\, M_{ij}$}   &   {$[O_{\!\mu}, O_{\!\nu}]  \longrightarrow i\, M_{\mu\nu}$}
     &  {$[O_{\!\ssc A}^{\prime} , O_{\!\ssc B}^{\prime} ]  \longrightarrow  i\, J_{\ssc \!A\!B}$} \\
\hline $\vec{u}^4=\frac{\gamma}{c}(c , v^i)  $ &
$\vec{\pi}^5=\frac{\Gamma}{\kappa\,c}(p^\mu , \kappa\,c) $  &
        $\vec{X}^6=\frac{G}{\ell}(z^{\ssc A} , \ell) $\\
{$\eta_{\mu\nu} u^\mu u^\nu = 1$}     &   {$\eta_{\ssc A\!B} \pi^{\ssc A} \pi^{\ssc B} = -1 $}
 &   {$\eta_{\ssc \mathcal M\mathcal N} X^{\!\ssc \mathcal M} X^{\!\ssc \mathcal N} = -1$}  \\
{$I\!\!R^3 \!\! \rightarrow SO(1,3)/SO(3)$}       &   {$I\!\!R^4  \!\!\rightarrow SO(1,4)/SO(1,3)$} &
   {$I\!\!R^5  \!\!\rightarrow SO(2,4)/SO(1,4)$} \\
\hline\hline
\end{tabular}
\end{center}\normalsize
\end{table}
%%%%%%%%%%%%%%%%%%%%%%%%%%%%%%%%%%%%%%%%%%%%%%%%%%%%%%

With 4D translations included, the Poincar\'e symmetry
{\small \boldmath $ISO(1,3)$ resulted is again unstable against
deformation. The stabilization is either {\small \boldmath $SO(1,4)$
or {\small \boldmath $SO(2,3)$. The Snyder suggestion of the Planck
mass  as a limiting energy-momentum  leads to {\small \boldmath
$SO(1,4)$, to be linearly realized on a 5D geometry\cite{023}. As
illustrated in the table,  $E^2-|\vec{p}|^2 c^2 \leq \kappa^2 c^4$
suggests the momentum five-vector $\vec{\pi}^5$ invariant under
{\small \boldmath $SO(1,4)$. Taking the bound as $|\vec{p}|^2 c^2
-E^2 \leq \kappa^2 c^4$ can be easily seen to give {\small \boldmath
$SO(2,3)$} instead. The latter case is obviously of no interest. The
new extra coordinate of the 5D geometry is to be identified as an
parameter $\sigma$ external to space-time giving, within {\small
\boldmath $ISO(1,3)$ picture, translations by $p^\mu\sigma$ ({\it c.f.}
$\Delta x=v^it$). The $\sigma$-coordinate has a spatial
geometric signature and a Einstein limit of proper time divided by
rest mass\cite{023}. It is neither space nor time.

It looks like the  5D geometry should also admits translational
symmetries.  The {\small \boldmath $ISO(1,4)$ resulted is again
mathematically unstable. On the other hand, having Planck mass as an
invariant may not be enough to get the relativity to describe a
quantum world.  In fact, without presuming the quantum $\hbar$, the
Planck length $\ell$ is an independent quantity. In the table, we also
show the deformation of {\small \boldmath $ISO(1,4)$  to {\small
\boldmath $SO(2,4)$} based on further imposing $\ell$ as an `length'
bound on the ultra-violet. The choice of {$|z^{\ssc A}|\leq
i\,\ell$} with $z^{\ssc A}$ as a `length' or `location' vector of
the 5D geometry gives a six-vector description of the `generalized
space-time location' $\vec{X}^6$ as an element of the AdS$_5$
geometry. Note that the $i$, explicitly, in {$|z^{\ssc A}|\leq
i\,\ell$} means
\[
(z^{\ssc 0})^2 - |\vec{z}|^2 -(z^{\ssc 4})^2 \leq -\ell^2 \;,
\]
hence an effective lower bound on {\it length} --- naively
$|\vec{z}|\geq \ell$ for $z^{\ssc 0}=z^{\ssc 4}$ (say both zero).
That is what is in line with the idea of a Planck length. To look at
the whole thing from this perspective, the complex $i$ for quantum
physics comes from the fact that the Planck scales as two, rather
than one, invariant quantities (like and beyond $c$) have to be
imposed as bounds on the part of the corresponding vectors with a
space-like and a  time-like signature, respectively. It is actually
the ``Minkowski" structure of the classical (six-) geometry that is
the true origin of the quantum $i$.

A strong advantage of having the last deformation achieved through
imposing a `length' bound is the fact that simple translations on the
resulted six-geometry cannot be admissible symmetries. That
terminates further extension to the unstable {\small \boldmath
$ISO(2,4)$} which will require further deformation according to the
philosophy behind the approach. Note that our extra dimensional
coordinates are not part of the space-time description at all at the
classical limit. Dynamics as behaviors concerning changes of spatial
locations or configurations with respect to time has apparently no
role for such coordinates. How to think about dynamics at the
{\small \boldmath $SO(2,4)$} invariant setting is a complete open
question at this point.

%\section
\noindent{\em The geometry and
 the scale/conformal transformation:-\ }
Similar to the case for dS$_5$ discussed in Ref.\cite{023}, the set of $z^{\ssc A}$'s simply
give a (Beltrami-type) five-coordinate description of the AdS$_5$ hypersurface
{$\eta_{\ssc \mathcal M\mathcal N} X^{\!\ssc \mathcal M} X^{\!\ssc \mathcal N} = -1$}.
In terms of $z^{\ssc A}$, the metric is given by
$g_{\ssc \!A\!B}= \frac{G^2}{{\ell}^2} \eta_{\ssc A\!B}
+\frac{G^4}{{\ell}^4} \eta_{\ssc A\!C}  \eta_{\ssc B\!D} z^{\ssc C} z^{\ssc D}$.
Introducing $q_{\ssc A} \equiv i\hbar \, \frac{\partial}{\partial z^{\ssc A}}$, and the
Lorentzian 5-vectors $Z_{\!\ssc  A}^{\ssc (\mathcal L)}
\equiv \eta_{\ssc A\!B} z^{\ssc B}= -\frac{\ell^2}{G^4}\,  z_{\ssc A}$ and
$P_{\!\ssc  A}^{\ssc (\mathcal L)}
%= \frac{1}{\ell} \left(x_{\!\ssc A}p_{\ssc 5}-x_{\ssc 5}p_{\!\ssc A}\right)
=q_{\ssc A} + Z_{\!\ssc  A}^{\ssc (\mathcal L)}\; \frac{1}{\ell^2}
\left(\eta^{\ssc B\!C} Z_{\!\ssc  B}^{\ssc (\mathcal L)}q_{\ssc C} \right)$, we have
representations of the {\small \boldmath $SO(2,4)$} generators given as
\be
J_{\!\ssc \mathcal M\mathcal N}
=  X_{\!\ssc \mathcal M}  P_{\!\ssc \mathcal N }
 - X_{\!\ssc \mathcal N}  P_{\!\ssc \mathcal M }
= Z_{\!\ssc  \mathcal M}^{\ssc (\mathcal L)} q_{\ssc  \mathcal N}
 -Z_{\!\ssc  \mathcal N}^{\ssc (\mathcal L)} q_{\ssc  \mathcal M}
= Z_{\!\ssc  \mathcal M}^{\ssc (\mathcal L)} P_{\!\ssc \mathcal N}^{\ssc (\mathcal L)}
 -Z_{\!\ssc  \mathcal N}^{\ssc (\mathcal L)} P_{\!\ssc  \mathcal M}^{\ssc (\mathcal L)} \;;
\ee
$ P_{\!\ssc \mathcal M }\equiv i \hbar \,\frac{\partial}{\partial X^{\ssc \mathcal M}}$, 
%= \frac{\ell}{G} q_{\ssc \mathcal M }$ and 
and we adopt the natural extended definitions   $Z_{\ssc  5}^{\ssc (\mathcal L)} \equiv  \ell$ and 
$P_{\!\ssc  5}^{\ssc (\mathcal L)}
%= \frac{1}{\ell} \left(x_{\!\ssc A}p_{\ssc 5}-x_{\ssc 5}p_{\!\ssc A}\right)
\equiv q_{\ssc 5} + Z_{\ssc  5}^{\ssc (\mathcal L)}\; \frac{1}{\ell^2}
\left(\eta^{\ssc B\!C} Z_{\!\ssc  B}^{\ssc (\mathcal L)}q_{\ssc C}
\right)=0$.

The `Lorentzian' 5-momentum $P_{\!\ssc  A}^{\ssc (\mathcal L)}=
-\frac{1}{\ell} J_{\!\ssc A5}$ is a quantum, noncommutative,
generalization of the `classical' 5-momentum at the level of the
intermediate {\small \boldmath $SO(1,4)$} relativity~\cite{023},
essentially as introduced by G\"{u}rsey~ \cite{Gur}. Moreover, its
first four components transform as that of a 4-vector under the 4D
Lorentz group {\small \boldmath $SO(1,3)$}. 
%The fifth component
%gives the operator from the deformation of central generator of the
%Heisenberg algebra :
%\[
%[\hat{X}_\mu,  \hat{P}_{\nu}] =- i \hbar \, \eta_{\mu\nu} \frac{1}{\kappa\,c\,\ell}
%J_{\!\ssc 45} = -  i \hbar \, \eta_{\mu\nu} \left(-\frac{1}{\kappa\,c}
%\, P_{\!\ssc  4}^{\ssc (\mathcal L)}\right) \;,
%\]
%where $\hat{P}_{\nu}=P_\nu^{\ssc (\mathcal L)}$ and $\hat{X}_\mu$
%identified as $-\frac{1}{\kappa\,c} J_{\!\mu\ssc 4}$. 
We also have
\be 
[ Z_{\!\ssc  A}^{\ssc (\mathcal L)}, P_{\!\ssc B}^{\ssc
(\mathcal L)} ] =-i \hbar \, \eta_{\ssc A\!B} \;. 
\ee 
Another note
worthy feature here is that $q_{\ssc 5}= -\frac{1}{\ell}\,
\left(\eta^{\ssc BC}Z_{\!\ssc B}^{\ssc (\mathcal L)}q_{\ssc C}
\right)$ resembles the conformal symmetry (scale transformation)
generator for the five-geometry with an otherwise Minkowski metric.
Translation along $z^{\ssc 5}\;(=\ell)$ is indeed a scaling of
$X^{\!\ssc \mathcal M}$. We explore another connection to 4D
conformal symmetry below.

%\section
%\noindent{\em Physics of the {\small \boldmath $SO(2,4)$} quantum relativity :-\ }
%\noindent{\em Connection to the scale/conformal transformation:-\ }
In the quantum regime, what one observes depends on the energy
scale the system is being probed. For high energy theorists, the
importance of the renormalization group evolutions cannot be
over-estimated. A quantum frame of reference will likely have to be
characterized also by the energy scale as the renormalization scale,
or some generalization of that. What is remarkable is that the
{\small \boldmath $SO(2,4)$} symmetry for the relativity is
mathematically the same group for conformal symmetry in 4D
space-time, usually considered as the symmetry for a scale invariant
theory. Our question here is how the relativity symmetry {\small
\boldmath $SO(2,4)$} can be connected to the 4D conformal symmetry
{\small \boldmath $SO(2,4)$}, and what that may teach us about the
physics of the Quantum Relativity.

Following Ref.\cite{023} (see also Refs.\cite{tsr,CO}) and discussion above,
we write our quantum  relativity algebra as :
\bea
&& [M_{\mu\nu}, M_{\lambda\rho}] =
i\hbar\, (\eta_{\nu\!\lambda} M_{\mu\rho} - \eta_{\mu\!\lambda} M_{\nu\rho}
+ \eta_{\mu\rho} M_{\nu\lambda} - \eta_{\nu\rho} M_{\mu\lambda})\;,
\nonumber\\
&& [M_{\mu\nu}, \hat{P}_{\!\lambda}] =  i \hbar \,(\eta_{\nu\!\lambda} \hat{P}_{\!\mu}
    - \eta_{\mu\!\lambda} \hat{P}_{\!\nu}) \;,
\nonumber \\
&& [M_{\mu\nu}, \hat{X}_{\!\lambda}] =  i \hbar \,(\eta_{\nu\!\lambda} \hat{X}_{\!\mu}
    - \eta_{\mu\!\lambda} \hat{X}_{\!\nu}) \;,
\nonumber \\
&& [\hat{X}_{\mu}, \hat{X}_{\!\nu}] =    \frac{i\hbar}{\kappa^2 c^2}  M_{\mu\nu} \;,
%\nonumber \\
\qquad
 [\hat{P}_{\mu}, \hat{P}_{\!\nu}] = -\,\frac{i\hbar}{\ell^2}  M_{\mu\nu} \;,
\nonumber \\
&& [\hat{X}_{\!\mu}, \hat{P}_{\!\nu}] = -i \hbar\, \eta_{\mu\nu} \hat{F} \;,
\qquad
 [\hat{X}_{\!\mu}, \hat{F}] = \frac{-i\hbar}{\kappa^2 c^2} \hat{P}_{\mu} \;,
\qquad [\hat{P}_{\!\mu}, \hat{F}] = \frac{-i \hbar}{\ell^2}
\hat{X}_{\mu}\;,
 \label{tsr1}
\eea
%where we have now shown $\hbar$ explicitly
($\hbar= {\kappa c \ell}$). This is to be matched to the standard form
\be \label{so} [J_{\ssc \!\mathcal R\mathcal S}, J_{\ssc\! \mathcal
M\mathcal N}] = i\hbar\, ( \eta_{\ssc \mathcal S\mathcal M} J_{\ssc
\mathcal R\mathcal N} - \eta_{\ssc \mathcal R\mathcal M} J_{\ssc
\mathcal S\mathcal N} + \eta_{\ssc \mathcal R\mathcal N} J_{\ssc
\mathcal S\mathcal M} -\eta_{\ssc \mathcal S\mathcal N} J_{\ssc
\mathcal R\mathcal M}) \;, %%
\ee
$J_{\ssc \!\mathcal M\mathcal N} = i\hbar\, (x_{\!\ssc \mathcal M}
\partial_{\!\ssc \mathcal N} -x_{\!\ssc  \mathcal N}\, \partial_{\!\ssc  \mathcal M})$.
%and  $\eta_{\ssc \mathcal M\mathcal N} =( 1, -1, -1, -1, -1, 1)$.
We identify
\bea
%%
%&& J_{\mu\nu} \equiv M_{\mu\nu} \;,
%\nonumber \\
&& J_{\mu \ssc 4} \equiv - {\kappa\, c} \; \hat{X}_{\mu}
=  i\hbar\, (x_{\mu}\partial_{\ssc 4}
 -x_{\ssc  4}\, \partial_{\mu}) \;,
\nonumber \\
&&  J_{\mu \ssc 5} \equiv -{\ell}\,\hat{P}_{\mu}
=  i\hbar\,  (x_{\mu}\partial_{\ssc 5}
 -x_{\ssc  5}\, \partial_{\mu})\;,
\nonumber \\
&& J_{\ssc 45}  \equiv {\kappa c \ell} \hat{F}=
 i\hbar\,  (x_{\ssc 4}\partial_{\ssc 5} -x_{\ssc  5}\, \partial_{\ssc 4}),
\quad J_{\mu\nu} \equiv M_{\mu\nu}.
\eea
The result gives an  interesting interpretation as
suggested by the notation that the generators represent a form of 4D
noncommutative geometry. The sets of $\hat{X}_{\!\mu}$'s and
$\hat{P}_{\!\mu}$'s give natural quantum generalizations of the
classical $x_\mu$'s and $p_\mu$'s (represented as $i\hbar
\partial_\mu$'s here), or $Z_{\!\mu}^{\ssc (\mathcal L)}$'s and
$q_\mu$'s in term of the five-geometry as discussed above. One can
check that they do have the right classical limit. Note that the algebra
may also be interpreted as coming from the stabilization of the
`Poincar\'e + Heisenberg' algebra with $\hat{F}$ being 
the central generator being deformation. On the AdS$_5$, 
$-{\kappa\, c}\,\hat{F}$ is $P_{\!\ssc  4}^{\ssc (\mathcal L)}$,
the fifth `momentum' component.

We introduce of the coordinates $x_+= (x_{\ssc 5}+x_{\ssc 4})/{\sqrt{2}}$
and $x_-=  (x_{\ssc 5}-x_{\ssc 4})/{\sqrt{2}}$, to be called conformal cone
coordinates.  The generators $J_{\mu\ssc 4}$ and $J_{\mu\ssc 5}$ may be
replaced by the equivalent set
\be
J_{\mu \pm} \equiv  i \hbar \, ( x_\mu\partial_\pm  - x_\pm \partial_{\mu})
= ( J_{\mu\ssc 5} \pm  J_{\mu\ssc 4}) /{\sqrt{2}}\;,
\ee
where
$\partial_\pm=  (\partial_{\ssc 5} \pm \partial_{\ssc 4})/{\sqrt{2}}$,
%\[ \frac{1}{\sqrt{2}}(x^5+x^4)=x^+=x_-=\frac{1}{\sqrt{2}}(x_5-x_4)
%\qquad \frac{1}{\sqrt{2}}(x^5-x^4)=x^-=x_+=\frac{1}{\sqrt{2}}(x_5+x_4) \]
and $J_{\ssc +-} \equiv i \hbar\, ( x_+ \partial_-  - x_- \partial_+) = J_{\ssc 45}$.
Mathematical structure of the algebra for conformal symmetry in 4D  Minkowski
space-time~\cite{adscft} can be obtained through the identification
\be\label{matching}
K_\mu \Rightarrow \sqrt{2}  J_{\mu -}\;,
%\nonumber \\
\qquad
P_\mu \Rightarrow \sqrt{2}  J_{\mu +}\;,
%\nonumber \\
\qquad
D \Rightarrow - J_{\ssc 45} \;.
\ee
% The conformal symmetry algebra  of {\small \boldmath $SO(2,4)$}
% for 4D space-time is given as
%%
% \bea &&
%  [M_{\mu\nu}, P_{\lambda}] =  i \hbar\,(\eta_{\nu\!\lambda} P_{\mu} - \eta_{\mu\!\lambda} P_{\nu}) \;, %ok
% \qquad
%  [M_{\mu\nu}, K_{\lambda}] =  i \hbar\,(\eta_{\nu\!\lambda} K_{\mu} - \eta_{\mu\!\lambda} K_{\nu}) \;,
% \nonumber \\
% &&
% [D, P_\mu] = - i\hbar\,  P_\mu \;, %ok
% \qquad\qquad\qquad
% [D, K_\mu] =  i\hbar\,  K_\mu \;,  %ok
% \nonumber \\
% &&
% [D, M_{\mu\nu}] = 0 \;,  %ok
% \qquad\qquad\qquad [P_\mu, K_\nu] = 2i\hbar\, (\eta_{\mu\nu} D
% -M_{\mu\nu}) \;,
% \nonumber \\
% &&
% [P_\mu, P_\nu] = 0  \;,  %ok
% \qquad\qquad\qquad
% [K_\mu, K_\nu] = 0  \;, %ok
% \label{conA}\eea
%%
%with $M_{\mu\nu}$ being the Lorentz transformation generators and
However, the physics picture is to be given by the definitions
\bea
&& P_\mu = i \hbar\, \partial_\mu^{\prime} \equiv  i \hbar\, \frac{\partial}{\partial y^\mu} \;,
\qquad\qquad D = i \hbar\, y^\mu  \partial_\mu^{\prime} \;,
\nonumber \\
&& K_\mu = i \hbar\, (2 y_\mu \, y^\nu  \partial_\nu^{\prime} - y^2 \,
\partial_\mu^{\prime} ) \;,
\label{conD}%%
\eea
where $y^\mu$ represents the  4-coordinate of Minkowski space-time.
%%%%%%%%%%%%%%%%%%%%
%%
% \be \label{jpk}
%J_{\mu\nu} = M_{\mu\nu} \;, \quad
% J_{\mu\ssc 4}= (P_\mu - K_\mu)\;, \quad J_{\mu\ssc 5}=
%  (P_\mu + K_\mu)\;, \quad J_{\ssc 45}=  -D \;.
% \ee
%%
% We want to find a sensible way to interpret the 4D conformal symmetry
% within the framework of the six-geometry above. Such a solution is as follows:
Recall that the introduction of the invariant length $\ell$ admits a
description of the coordinate variable $x$ as a pure number (denoted
rather by $X$ above). Obviously, the standard 6-coordinate
definition for $J_{\ssc \!\mathcal M\mathcal N}$ is invariant under
such re-scaling.  Next, we consider the 6- to 4- coordinate
transformation on a special 4D hypersurface to be given by the
dimensionless $(x^\mu, x^{\ssc 4}, x^{\ssc 5})= (y^\mu, \frac{1}{2}
\eta_{\mu\nu} y^\mu y^\nu +\frac{1}{2}, \frac{1}{2} \eta_{\mu\nu}
y^\mu y^\nu -\frac{1}{2})$.
%  \[ e^\mu_\nu =\delta^\mu_\nu   \qquad e^{\ssc 4}_\mu =e^{\ssc 4}_\mu =\eta_{\mu\nu} y^\nu
%     \qquad g_{\mu\!\nu}=  \eta_{\ssc \!\mathcal M\mathcal N} e^{\ssc \!\mathcal M}_\mu e^{\ssc \!\mathcal N}_\nu  =  \eta_{\mu\!\nu} \]
%  \[ E^\mu_\nu =\delta^\mu_\nu
%     \qquad E_{\ssc 5}^\mu =  \eta_{\ssc 5 A} g^{\mu\!\nu}  e^{\ssc 5}_\nu  =  \eta_{\ssc 55} g^{\mu\!\nu}  e^{\ssc 5}_\nu    = -\eta^{\mu\nu} \eta_{\nu\lambda} y^\lambda = y^\mu =-E_{\ssc 4}^\mu     \]
One easily sees that the metric in terms of $y^\mu$ is still $\eta_{\mu\!\nu}$, hence
Minkowski. Moreover, we have
\bea \label{cu}
 && x_+=x^-%=\frac{1}{\sqrt{2}} (x^{\ssc 5} -x^{\ssc 4})
= -1/{\sqrt{2}}  \;,
\qquad  x_- =x^+= y^2/{\sqrt{2}}  \;,
\nonumber \\
 && \partial_+ = 0  \;, \qquad
 \partial_{\ssc 5} =-\partial_{\ssc 4}
=\frac{1}{\sqrt{2}}  \,  \partial_-
=x^\nu  \partial_\nu  \;.
\eea
The latter does give exactly Eq.(\ref{conD}) through expression (\ref{matching}).
%%
%\be \label{x45} x_{\ssc 4}= - {1\over 2} \,
%x^2 - {1\over 2} \;, \qquad x_{\ssc 5}=  {1\over 2} \,   x^2  -{1\over 2} \; \ee
%%
%For instance,  we have then $\sqrt{2}  J_{\mu +}$  reducing,
% explicitly, to $ i \hbar\,  \partial_{\mu}$; and
%%
%\be [\partial_-, \partial_{\mu}] = -2 \, \partial_{\mu} \;, \qquad
%[\partial_-, x_{\mu}] = 2 \, x_{\mu} \;, \qquad [\partial_-, x_-] =4 \, x^2 \;. \ee
So, we can say that for the 4D hypersurface in the six-geometry satisfying
Eq.(\ref{cu}), translations along $x_{\ssc 4}$ and $x_{\ssc 5}$ do correspond to
scaling, as $i\hbar\,\partial{\ssc 5} =-i\hbar\,\partial_{\ssc 4}=D$ . We call this
hypersurface the conformal universe. The latter satisfies
$\eta_{\ssc \mathcal M\!N} x^{\ssc \mathcal M} x^{\ssc \mathcal N}=0$
%%
%\be
%\eta_{\ssc \mathcal M\!N} x^{\ssc \mathcal M} x^{\ssc \mathcal N}
%=  \eta_{\mu\!\nu} x^\mu x^nu +2\, x_+\, x_- = 0 \; ,
%\ee
%%
while the quantum world has $\eta_{\ssc \mathcal M\!N} x^{\ssc \mathcal M}
 x^{\ssc \mathcal N}=-1$ as shown in Table~1. The $J_{\ssc \!\mathcal M\mathcal N}$
transformations of {\small \boldmath $SO(2,4)$} leaves both invariant.
We are then forced to conclude that the quantum world cannot have 4D
scale invariance. The analysis also illustrates that translations along $x_{\ssc 4}$
and $x_{\ssc 5}$ can be considered as some sort of scaling, or transformation
(energy) scale. To establish the latter on a more solid setting, we do need to
first build a theory of dynamics, which is beyond the scope of the present letter.

\noindent{\em Conclusion :-\ } Special Einstein relativity as given
by {\small \boldmath $SO(1,3)$} is the deformation or stabilization
of the Galilean {\small \boldmath $ISO(3)$}. Along the same line,
extending to {\small \boldmath $ISO(1,3)$} and stabilizing to
{\small \boldmath $SO(1,4)$} has been considered as admitting the
deforming parameter $\frac{1}{\kappa^2c^2}$ to be nonzero. While
this gives the finite Planck mass $\kappa c$, there is still
no $\hbar$. Going further to {\small \boldmath $ISO(1,4)$} and then
{\small \boldmath $SO(2,4)$} may be taken as admitting independently
the finite Planck length $\ell$. The latter together with $\kappa\,
c$ gives $\hbar$.  The symmetry for quantum relativity is hence
{\small \boldmath $SO(2,4)$}, the linear realization of which tells
that the quantum world is an AdS$_5$ sitting inside a classical
six-geometry of four space-time plus two extra coordinates. {\small
\boldmath $ISO(2,4)$} is not a symmetry for the AdS$_5$, hence no
further extension and deformation. The formulation also gives a quantum,
noncommutative, 4D space-time description, fitting well with the natural
perspective from the deformation approach that these extra coordinates
are neither space nor time. They are connected to the concept of
(energy) scale, though the quantum world is not scale invariant but
rather `parallel' to the conformal universe.

The relativity symmetry stabilization approach, with the quite
minimal physics input of having the fundamental constants Planck
mass and Planck length (hence also the quantum $\hbar$) being the
deformation parameters is illustrated to give an AdS$_5$ as the
quantum world with {\small \boldmath $SO(2,4)$} as the reference
frame transformation symmetry. That is but all kinematics, the next
challenge is to build a theory of dynamics, or a theory that does
give us dynamics as we know it at the classical space-time limit.
The latter represents further big
challenges to our conceptual thinking about fundamental physics.

%\acknowledgements
\noindent{\em Acknowledgements :-\ }
%%
%Otto Kong would like to thank C.-M.~Chen and F.-L.~Lin for helpful discussions.
The author is deeply in debt to X.~Tata,  discussions and questions from whom
inspired the present work. This work is partially supported by the research Grant No.
95-2112-M-008-001 from the NSC of Taiwan.

%%%%%%%%%%%%%%%%%

\end{document}